\documentclass[showpacs,preprintnumbers,amsmath,amssymb,nofootinbib,floatfix]{revtex4}

\usepackage[dvips]{graphicx}
\usepackage{dcolumn}
\usepackage{bm}

\def\mapgeq{\mathbin{\lower.3ex\hbox{$\buildrel>\over{\smash{\scriptstyle\sim}\vphantom{_x}}$
}}}
\def\mapleq{\mathbin{\lower.3ex\hbox{$\buildrel<\over{\smash{\scriptstyle\sim}\vphantom{_x}}$
}}}
\def\mapgeqeq{\mathbi{\lower.3ex\hbox{$\buildrel>\over{\smash{\scriptstyle\approx}\vphantom{_2}}$
}}}
\def\mapleqeq{\mathbin{\lower.3ex\hbox{$\buildrel<\over{\smash{\scriptstyle\approx}\vphantom{_2}}$
}}}
\mathchardef\hanaO="724F
\def\Journal#1#2#3#4{{#1} {\bf #2}, #3 (#4)}

\def\NPB{Nucl. Phys. B}

\def\NPSUPPL{Nucl. Phys. Proc. Suppl.}
\def\PLB{{Phys. Lett.} B}

\def\PLBOLD{Phys. Lett.}
\def\PRL{Phys. Rev. Lett.}
\def\RMP{Rev. Mod. Phys.}
\def\PRD{Phys. Rev. D}

\def\PTP{Prog. Theor. Phys.}
\def\JHEP{JHEP}

\def\EPJ{Euro. Phys. J. C}

\def\JETPUSSR{Sov. Phys. JETP}

\def\ZETP{Zh. Eksp. Teor. Fiz.}

\def\IJMP{Int. J. Mod. Phys. A}

\def\JPG{J. Phys. G}

\def\SCI{Science}
\def\APJ{Astrophysics J.}

\def\NJP{New. J. Phys.}

\def\Erratum{Erratum-ibid}

\begin{document}

\preprint{TOKAI-HEP/TH-0601}

 \title{What Does $\mu$-$\tau$ Symmetry Imply about Neutrino Mixings?}

\author{Kenichi Fuki}
\email{fuki@phys.metro-u.ac.jp}
\affiliation{\vspace{3mm}%
\sl Department of Physics, Tokyo Metropolitan University,\\
1-1 Minami-Osawa, Hachioji, Tokyo 192-0397, Japan
}
\author{Masaki Yasu\`{e}}%
\email{yasue@keyaki.cc.u-tokai.ac.jp}
\affiliation{\vspace{3mm}%
\sl Department of Physics, Tokai University,\\
1117 Kitakaname, Hiratsuka, Kanagawa 259-1292, Japan
}

\date{January, 2006}

\begin{abstract}
The requirement of the $\mu$-$\tau$ symmetry in the neutrino sector that yields 
the maximal atmospheric neutrino mixing 
is shown to yield either $\sin\theta_{13}=0$ (referred to as C1)) or $\sin\theta_{12}=0$
 (referred to as C2)), 
where $\theta_{12(13)}$ stands for the solar (reactor) neutrino mixing angle.  
We study general properties possessed by 
approximately $\mu$-$\tau$ symmetric textures.  It is argued that the tiny $\mu$
-$\tau$ symmetry breaking generally leads 
to $\cos 2\theta_{23}\sim\sin\theta_{13}$ for C1) and 
$\cos 2\theta_{23}\sim \Delta m^2_\odot/\Delta m^2_{atm}(\equiv R)$ 
for C2), which indicates that the smallness of $\cos 2\theta_{23}$ is a good 
measure of the $\mu$-$\tau$ symmetry 
breaking, where $\Delta m^2_{atm}$ ($\Delta m^2_\odot$) stands for the square 
mass differences of atmospheric (solar) 
neutrinos.  We further find that the relation $R\sim\sin^2\theta_{13}$ arises 
from contributions of 
${\mathcal{O}}(\sin^2\theta_{13})$ in the estimation of the neutrino masses (
$m_{1,2,3}$) for C1), and that possible 
forms of textures are strongly restricted to realize $\sin^22\theta_{12}={\mathcal{O}}(1)$
 for C2).  To satisfy 
$R\sim\sin^2\theta_{13}$ for C1), neutrinos exhibit the inverted mass hierarchy,
 or the quasi degenerate mass pattern 
with $\vert m_{1,2,3}\vert \sim {\mathcal{O}}(\sqrt{\Delta m^2_{atm}})$, and, to realize 
$\sin^22\theta_{12}={\mathcal{O}}(1)$ for C2), there should be an additional small 
parameter $\eta$ whose size is 
comparable to that of the $\mu$-$\tau$ symmetry breaking parameter $\varepsilon$, giving 
$\tan 2\theta_{12} \sim \varepsilon/\eta$ with $\eta\sim \varepsilon$ to be 
compatible with the observed large mixing.

\end{abstract}

\maketitle
\section{\label{sec:1}Introduction}
Since the confirmation of the atmospheric neutrino oscillations in 1998 \cite{SK}, 
neutrino oscillations have been observed in various neutrinos \cite{Experiments}
 coming from the Sun \cite{OldSolor,Sun}, accelerators \cite{K2K} and  reactors 
\cite{Reactor}.  The results of the neutrino oscillations are known to be interpreted 
in terms of the mixings of three flavor neutrinos, $\nu_e$, $\nu_\mu$ and $\nu_\tau$, 
which evolve into three massive neutrinos $\nu_1$, $\nu_2$ and $\nu_3$ during 
their flights.  Observed in experiments are three mixing angles denoted by 
$\theta_{12}$ for $\nu_e$-$\nu_\mu$, $\theta_{23}$ for $\nu_\mu$-$\nu_\tau$ and 
$\theta_{13}$ for $\nu_e$-$\nu_\tau$ and two neutrino mass squared differences 
$\Delta m^2_{atm}$ for atmospheric neutrinos and $\Delta m^2_\odot$ for solar 
neutrinos.  These masses and mixing angles are currently constrained to satisfy 
\cite{NuData}:
\begin{eqnarray}
\Delta m^2_\odot = 7.92\left( 1\pm0.09\right) \times 10^{-5}~{\rm eV}^2,
\quad
\Delta m^2_{atm} = 2.4
{\footnotesize
\left( {1\begin{array}{*{20}c}  
 { + 0.21}  \\  
 { - 0.26}  \\
\end{array}} \right)
} \times 10^{-3}~{\rm eV}^2,
\label{Eq:NuDataMass}
\end{eqnarray}
and
\begin{eqnarray}
\sin ^2 \theta _{12}  = 0.314
{\footnotesize
\left( {1\begin{array}{*{20}c}  
 { + 0.18}  \\  
 { - 0.15}  \\
\end{array}} \right)}
,
\quad
\sin ^2 \theta _{23}  = 0.44
{\footnotesize
\left( {1\begin{array}{*{20}c}  
 { + 0.41}  \\  
 { - 0.22}  \\
\end{array}} \right),
}
\quad
\sin ^2 \theta _{13}  = 0.9 
{\footnotesize
\left( {\begin{array}{*{20}c}  
 { + 2.3}  \\  
 { - 0.9}  \\
\end{array}} \right) \times 10^{-2},
}
\label{Eq:NuDataAngle}
\end{eqnarray}
where $\Delta m^2_\odot = m^2_2-m^2_1$ $(>0)$ \cite{PositiveSolor} and 
$\Delta m^2_{atm} = \vert m^2_3-(m^2_1+m^2_2)\vert/2$ and $m_1$, $m_2$ and $m_3$, 
respectively, stand for the masses of $\nu_1$, $\nu_2$ and $\nu_3$. These 
experimental data have indicated two distinct properties: 1) The atmospheric and 
solar mixing angles measured as $\sin^2 2\theta$ are ${\mathcal{O}}(1)$ while the 
reactor mixing angle $\theta_{13}$ is quite small, and 2) The mass squared 
differences $\Delta m^2_{atm}$ and $\Delta m^2_\odot$ exhibit the hierarchy as 
$\Delta m^2_\odot/\Delta m^2_{atm}\ll 1$.

It has been a guiding principle that the presence of hierarchies or of tiny 
quantities implies a presence of a certain protection symmetry in underlying 
physics \cite{tHooft}.  Candidates of such a symmetry in neutrino physics 
\cite{NuSymmetry} may include $U(1)_{L^\prime}$ based on the conservation of 
$L_e-L_\mu-L_\tau$ ($\equiv L^\prime$) \cite{EarlierLprime,Lprime}, and a $\mu$
-$\tau$ symmetry based on the invariance of flavor neutrino mass terms under the 
interchange of $\nu_\mu$ and $\nu_\tau$ characterized by $Z_2$ 
\cite{Nishiura,mu-tau,mu-tau0,mu-tau1,mu-tau2}, 
where $L_{e,\mu,\tau}$, respectively, represent the $e$-, $\mu$-, and $\tau$
-number.  These symmetries show that $\Delta m^2_\odot/\Delta m^2_{atm}=0$, 
$\sin^2 2\theta_{12}=1$ and $\sin\theta_{13}=0$ for $U(1)_{L^\prime}$ and 
$\sin^2 2\theta_{23}=1$ and $\sin\theta_{13}=0$ for the $\mu$-$\tau$ symmetry.  
Since the charged leptons clearly violate these symmetries, the effect from the 
charged leptons yields deviations from these values and we expect that their 
contributions finally give compatible results with Eqs.(\ref{Eq:NuDataMass}) and 
(\ref{Eq:NuDataAngle}). However, the charged leptons and neutrinos are the $SU(2)_L$-doublets 
and the $\mu$-$\tau$ symmetry respected by neutrinos should be respected by the
charged leptons.  This fact apparently disfavors the requirement of the $\mu$-$\tau$ symmetry.
To have $\mu$-$\tau$ symmetric mass terms, we must introduce several Higgs scalars with the 
different $Z_2$ parity, where their vev's can provide charged lepton masses 
\cite{mu-tau0,mu-tau2,leptonic-mu-tau} in such a way that the charged leptons acquire 
almost diagonal masses, which badly break the $\mu$-$\tau$ symmetry. The price to pay 
is to have flavor-changing neutral current interactions due to the direct exchanges 
of these Higgs scalars.  Effects from the interactions become sizable for quraks when 
the $\mu$-$\tau$ symmetry is applied to grand unified models \cite{unified-mu-tau} and 
should be suppressed. If neutrinos are Majorana particles, which are different from 
charged leptons of the Dirac type, it is expected that this difference may 
supply approximately $\mu$-$\tau$ symmetric neutrino flavor structure. 

In this article, we discuss details of physical results from the requirement of 
the $\mu$-$\tau$ symmetry in neutrino mixings without CP phases.  The influence 
from CP phases will be discussed in a subsequent article \cite{NewCP}.  Throughout 
this article, we assume that the effects from the charged leptons are fully 
contained in our discussions as $\mu$-$\tau$ symmetry breaking effects. We calculate 
eigenvectors associated with a given flavor neutrino mass matrix $M_\nu$, which 
determines the Pontecorvo-Maki-Nakagawa-Sakata unitary matrix $U_{PMNS}$ \cite{PMNS}
 that converts the flavor neutrinos $\nu_{e,\mu,\tau}$ into the massive neutrinos 
$\nu_{1,2,3}$: $\nu_f = (U_{PMNS})_{fi}\nu_i$, where $f$=$e,\mu,\tau$ and $i$=1,
2,3. The genuine use of the $\mu$-$\tau$ symmetry indicates two categories of 
$\mu$-$\tau$ symmetric textures, which give either $\sin\theta_{13}=0$, or 
$\sin\theta_{12}=0$ in $U_{PMNS}$ depending on the order of the eigenvalues.  
After including a $\mu$-$\tau$ symmetry breaking effect characterized by a parameter 
$\varepsilon$, yielding either $\sin\theta_{13}\sim\varepsilon$, or 
$\sin\theta_{12}\sim\varepsilon$, we find general constraints on flavor neutrino 
masses that yield $\sin^3\theta_{13}\ll 1$ and $\Delta m^2_\odot/\Delta m^2_{atm} \ll 1$.  
Furthermore, to obtain $\sin^22\theta_{12}={\mathcal{O}}(1)$ from 
$\sin\theta_{12}\sim \varepsilon$ gives a severe constraint on sizes of the flavor 
neutrino masses.  The $\mu$-$\tau$ symmetry breaking results in a correlation of 
$\cos\theta_{23}$ to $\sin\theta_{13}$, or to $\Delta m^2_\odot/\Delta m^2_{atm}$
 \cite{Theta31AndMass}.  If $m_1+m_2\sim 0$ giving $\Delta m^2_\odot\sim 0$, we 
show that the relation of $\Delta m^2_\odot/\Delta m^2_{atm} \sim \sin^2\theta_{13}$
 arises in textures leading to $\sin\theta_{13}=0$ in the $\mu$-$\tau$ symmetric 
limit.

In the next section, we explain how $\sin\theta_{13}=0$ and $\sin\theta_{12}=0$ 
are obtained in the $\mu$-$\tau$ symmetric limit.  The useful formula are shown 
to calculate neutrino masses and mixing angles, where two categories depend on 
the signs of $\sin\theta_{23}$ for a given $M_\nu$.  In Sec.\ref{sec:3}, we derive 
various constraints to realize $\sin^2\theta_{13}\ll 1$ and $\sin^22\theta_{12}={\mathcal{O}}(1)$.  
In Sec.\ref{sec:4}, applying these constraints to textures, we find general 
relations among $\cos 2\theta_{23}$, $\sin^2\theta_{13}$, and $\Delta m^2_\odot/\Delta m^2_{atm}$, 
which do not depend on details of textures.\footnote{Specific forms of textures 
that respect our constraints will be presented elsewhere to make testable predictions 
\cite{New}.}  The final section, Sec.\ref{sec:5}, is devoted to summary, and 
discussions.

\section{\label{sec:2}$\mu$-$\tau$ Symmetric Texture}
Let us define a neutrino mass matrix $M_\nu$ parameterized by\footnote{It is 
understood that the charged leptons and neutrinos are rotated, if necessary, to 
give diagonal charged-current interactions and to define the flavor neutrinos of 
$\nu_e$, $\nu_\mu$ and $\nu_\tau$.}
\begin{eqnarray}
&& M_\nu = \left( {\begin{array}{*{20}c}
	M_{ee} & M_{e\mu} & M_{e\tau}  \\
	M_{e\mu} & M_{\mu\mu} & M_{\mu\tau}  \\
	M_{e\tau} & M_{\mu\tau} & M_{\tau\tau}  \\
\end{array}} \right).
\label{Eq:NuMatrixEntries}
\end{eqnarray}
The $\mu$-$\tau$ symmetry is based on the invariance of the flavor neutrino mass 
terms in the lagrangian under the interchange of $\nu_\mu \leftrightarrow \nu_\tau$
 or $\nu_\mu \leftrightarrow -\nu_\tau$.  As a result, we obtain $M_{e\tau}=M_{e\mu}$
 and $M_{\mu\mu}=M_{\tau\tau}$ for $\nu_\mu \leftrightarrow \nu_\tau$ or 
$M_{e\tau}=-M_{e\mu}$ and $M_{\mu\mu}=M_{\tau\tau}$ for $\nu_\mu \leftrightarrow -\nu_\tau$.  
We use the sign factor $\sigma=\pm 1$  to have $M_{e\tau}=-\sigma M_{e\mu}$ 
for the $\mu$-$\tau$ symmetric part under the interchange of 
$\nu_\mu\leftrightarrow -\sigma\nu_\tau$.  We divide $M_\nu$ into the $\mu$-$\tau$
 symmetric part $M_{sym}$ and its breaking part $M_b$ \cite{mu-tau-breaking,MassTextureCP,MassTextureCP1}
 expressed in terms of $M^{(\pm)}_{e\mu} = (M_{e\mu} \pm (-\sigma M_{e\tau}))/2$
 and $M^{(\pm)}_{\mu\mu} = (M_{\mu\mu} \pm M_{\tau\tau})/2$:
\begin{eqnarray}
&&
M_\nu = M_{sym} + M_b
\label{Eq:Mnu-mutau-separation}
\end{eqnarray}
with
\begin{eqnarray}
&&
M_{sym}  = \left( \begin{array}{*{20}c}  
 M_{ee} & M^{(+)}_{e\mu } & - \sigma M^{(+)}_{e\mu }  \\  
 M^{(+)}_{e\mu } & M^{(+)}_{\mu\mu } & M_{\mu \tau }   \\  
  - \sigma M^{(+)}_{e\mu } & M_{\mu \tau } & M^{(+)}_{\mu\mu }\\
\end{array} \right),
\quad
M_b  = \left( \begin{array}{*{20}c}  
 0 & M^{(-)}_{e\mu } & \sigma M^{(-)}_{e\mu }  \\  
 M^{(-)}_{e\mu }& M^{(-)}_{\mu\mu } & 0  \\  
 \sigma M^{(-)}_{e\mu } & 0 & - M^{(-)}_{\mu\mu } \\
\end{array} \right),
\label{Eq:Mnu-mutau-separation-2}
\end{eqnarray}
where obvious relations of $M_{e\mu}=M^{(+)}_{e\mu}+M^{(-)}_{e\mu}$, 
$M_{e\tau}=- \sigma(M^{(+)}_{e\mu}-M^{(-)}_{e\mu})$, $M_{\mu\mu}=M^{(+)}_{\mu\mu}+M^{(-)}_{\mu\mu}$
 and $M_{\tau\tau}= M^{(+)}_{\mu\mu}-M^{(-)}_{\mu\mu}$ are used.  The lagrangian 
for $M_{sym}$: $-{\mathcal{L}}_{mass}=\psi^TM_{sym}\psi/2$ with 
$\psi=(\nu_e, \nu_\mu, \nu_\tau)^T$ turns out to be invariant under the exchange 
of $\nu_\mu\leftrightarrow -\sigma\nu_\tau$. 

It is not difficult to find three eigenvalues of the $\mu$-$\tau$ symmetric 
$M_{sym}$.  After a little calculus, we obtain three eigenvalues $\lambda_\pm$ 
and $\lambda$ 
\begin{eqnarray}
&&
\lambda _ \pm   = M^{(+)}_{\mu \mu }  - \sigma M_{\mu \tau }  + M^{(+)}_{e\mu } x_ \pm,  
\quad
\lambda  = M^{(+)}_{\mu \mu }  + \sigma M_{\mu\tau},
\label{Eq:EigenValues}
\end{eqnarray}
where
\begin{eqnarray}
&&
x_ \pm   = \frac{{M_{ee}  - M^{(+)}_{\mu \mu }  + \sigma M_{\mu\tau}  \pm \sqrt 
{\left( {M_{ee}  - M^{(+)}_{\mu \mu }  + \sigma M_{\mu\tau} } \right)^2  + 8M^{
(+)2}_{e\mu } } }}{{2M^{(+)}_{e\mu } }}.
\label{Eq:x-pm}
\end{eqnarray}
The ordering of $\vert\lambda_\pm\vert$ and $\vert\lambda\vert$ determines masses 
of $\nu_{1,2,3}$.  For example, if $\vert\lambda_-\vert<\vert\lambda_+\vert<\vert\lambda\vert$, 
the neutrino masses are given by $m_1=\lambda_-$, $m_2=\lambda_+$ and $m_3= \lambda$
 as the normal mass hierarchy and by $m_1=\lambda_+$, $m_2=\lambda$ and $m_3=\lambda_-$
 as the inverted mass hierarchy.  The quasi degenerate mass pattern further 
requires $\vert m_i-m_j\vert \ll \vert m_{1,2,3}\vert$ ($i,j=1,2,3$).  These two 
examples show the typical cases, where $\lambda$ is assigned to $\nu_3$ or to the 
others.  We will see that, if $\lambda$ is assigned to the  mass of $\nu_3$, 
$\sin\theta_{13}=0$ is derived, while if $\lambda$ is assigned to the mass of 
$\nu_2$, $\sin\theta_{12}=0$ is derived.\footnote{This kind of consideration has 
been done in Ref.\cite{Nishiura2}, where it was mainly applied to the analysis 
on the CKM unitary matrix \cite{CKM}.  See also Ref.\cite{mu-tau-breaking}, where 
the possible choice of $\sin\theta_{12}=0$ was phrased as the requirement of 
$m_1=m_2$, whose consequence was not fully discussed.} The eigenvectors are also 
calculated to be
\begin{eqnarray}
&&
\left| \lambda_- \right\rangle  = n_- \left( {\begin{array}{*{20}c}  
 {-x_ - }  \\  
 -1  \\  
 { \sigma }  \\
\end{array}} \right),
\quad
\left| \lambda_+ \right\rangle  = n_+ \left( {\begin{array}{*{20}c}  
 {x_ + }  \\  
 1  \\  
 { - \sigma }  \\
\end{array}} \right),
\quad
\left| \lambda \right\rangle  = \frac{1}{{\sqrt 2 }}\left( {\begin{array}{*{20}c}  
 0  \\  
 \sigma   \\  
 1  \\
\end{array}} \right),
\label{Eq:EigenVectors}
\end{eqnarray}
respectively, for $\lambda_-$, $\lambda_+$ and $\lambda$, where $n_\pm = \sqrt{2 + x^2_\pm}$.  
These eigenvectors are determined up to an arbitrary relative phase as long 
as corresponding eigenvalues remain intact.  The orthogonally condition is obviously 
satisfied because of $x_+x_-=-2$.  These eigenvectors form the PMNS unitary matrix 
expressed by
\begin{eqnarray}
U^{(0)}_{PMNS} &=&  
  \left(   
  \begin{array}{ccc}  
  c_{12}c_{13}                     &  s_{12}c_{13}   &  s_{13}\\  
  -s_{12}c_{23}-c_{12}s_{23}s_{13} &  c_{12}c_{23}-s_{12}s_{23}s_{13}  &  s_{23}c_{13}\\  
  s_{12}s_{23}-c_{12}c_{23}s_{13}  &  -c_{12}s_{23}-s_{12}c_{23}s_{13} & c_{23}c_{13}\\  
  \end{array}   
  \right),
\label{Eq:PMNS-noCP}
\end{eqnarray}
where $c_{ij} \equiv \cos\vartheta_{ij}$ and $s_{ij} \equiv \sin\vartheta_{ij}$.

We first assume that $M_\nu$ gives $\vert\lambda_-\vert<\vert\lambda_+\vert<\vert\lambda\vert$
 in the normal mass hierarchy and, thus, assign $\lambda$ to $m_3$.  As a result,
 $U_{PMNS}$ can be described by
\begin{eqnarray}
&&
\left( {\frac{1}{{\sqrt {2 + \left( {x_ -  } \right)^2 } }}\left( {\begin{array}{*{20}c}  
 { - x_ -  }  \\  
 { - 1}  \\  
 \sigma   \\
\end{array}} \right) \frac{1}{{\sqrt {2 + \left( {x_ +  } \right)^2 } }}\left( {
\begin{array}{*{20}c}  
 {x_ +  }  \\  
 1  \\  
 { - \sigma }  \\
\end{array}} \right) \frac{1}{{\sqrt 2 }}\left( {\begin{array}{*{20}c}  
 0  \\  
 \sigma   \\  
 1  \\
\end{array}} \right)} \right).
\label{Eq:U_PMNS_13case}
\end{eqnarray}
By comparing it with $U^{(0)}_{PMNS}$, we obtain that
\begin{eqnarray}
&&
\tan 2\theta _{12}  = \frac{{2\sqrt 2 M_{e\mu }^{( + )} }}{{M_{\mu \mu }^{( + )}  
- \sigma M_{\mu \tau } } - M_{ee}},
\quad
\sin 2\theta _{23}  =\sigma,
\quad
\sin \theta _{13}  =0.
\label{Eq:mixing_13case}
\end{eqnarray}
This prediction leads to the statement that the $\mu$-$\tau$ symmetry guarantees 
the appearance of the maximal atmospheric neutrino mixing, and of the vanishing 
$\theta_{13}$.  The $\mu$-$\tau$ symmetric mass matrix with 
$\vert\lambda_-\vert<\vert\lambda_+\vert<\vert\lambda\vert$ itself can give 
consistent mixing angles with the experimental data provided that an appropriate 
magnitude of $\theta_{12}$ is produced.  The similar conclusion can be obtained 
for $\vert\lambda\vert<\vert\lambda_-\vert<\vert\lambda_+\vert$ in the inverted 
mass hierarchy, where $m_3$ is still assigned to $\lambda$.

Starting with the same $M_\nu$, we next assume $\vert\lambda_+\vert<\vert\lambda\vert<\vert\lambda_-\vert$
 in the normal mass hierarchy, and assign $\lambda$ to $m_2$.  In this mass-ordering,
 we construct $U_{PMNS}$ to be:
\begin{eqnarray}
&&
\left( {\frac{1}{{\sqrt {2 + \left( {x_ +  } \right)^2 } }}\left( {\begin{array}{*{20}c}  
 {x_ +  }  \\  
 -1  \\  
 { \sigma }  \\
\end{array}} \right)\frac{1}{{\sqrt 2 }} \left( {\begin{array}{*{20}c}  
 0  \\  
 1   \\  
 \sigma  \\
\end{array}} \right) \frac{1}{{\sqrt {2 + \left( {x_ -  } \right)^2 } }}\left( {
\begin{array}{*{20}c}  
 { \sigma x_ -  }  \\  
 { - \sigma}  \\  
 1  \\
\end{array}} \right)} \right),
\label{Eq:U_PMNS_12case}
\end{eqnarray}
from which  we obtain that
\begin{eqnarray}
&&
\sin\theta _{12}  = 0,
\quad
\sin 2\theta _{23}  = -\sigma,
\quad
\tan 2\theta _{13}  = -\frac{{2\sqrt 2 \sigma M_{e\mu }^{( + )} }}{{M_{\mu \mu }
^{( + )}  - \sigma M_{\mu \tau } } - M_{ee}},
\label{Eq:mixing_12case}
\end{eqnarray}
Therefore, we have $\sin\theta_{12}=0$ instead of $\sin\theta_{13}=0$.  The similar 
conclusion can be obtained for $\vert\lambda_-\vert<\vert\lambda_+\vert<\vert\lambda\vert$
 in the inverted mass hierarchy, where $m_2$ is still assigned to $\lambda$.

There is a general formula \cite{FormulaNoCP} that can treat both cases in a 
unified way.  The mixing angles and masses are given by
\begin{eqnarray}
&&
\tan 2\theta _{12} = \frac{2X}{\lambda _2  - \lambda _1},
\nonumber\\
&&
\left( M_{\tau\tau} - M_{\mu\mu}\right)\sin 2\theta_{23}  - 2 M_{\mu\tau}\cos 2
\theta_{23}= 2s_{13} X,
\nonumber\\
&&
\tan 2\theta _{13}  = \frac{2 Y}{\lambda _3 - M_{ee}},
\label{Eq:ExactMixingAngles}
\end{eqnarray}
and
\begin{eqnarray}
&&
m_1  = c_{12}^2 \lambda_1  + s_{12}^2 \lambda _2  - 2c_{12} s_{12} X,
\quad
m_2 = s_{12}^2 \lambda_1  + c_{12}^2 \lambda _2  + 2c_{12} s_{12} X,
\nonumber\\
&&
m_3  = c_{13}^2 \lambda _3  + 2c_{13} s_{13}Y + s_{13}^2 M_{ee},
\label{Eq:ExactMasses}
\end{eqnarray}
where
\begin{eqnarray}
&&
X = \frac{c_{23} M_{e\mu} - s_{23} M_{e\tau}}{c_{13}},
\quad
Y = s_{23} M_{e\mu} + c_{23} M_{e\tau},
\nonumber\\
&&
\lambda_1  = c_{13}^2 M_{ee} - 2c_{13} s_{13} Y + s_{13}^2 \lambda_3,
\quad
\lambda_2  = c_{23}^2 M_{\mu\mu} + s_{23}^2 M_{\tau\tau} - 2s_{23} c_{23} M_{\mu\tau},
\nonumber\\
&&
\lambda_3  = s_{23}^2 M_{\mu\mu} + c_{23}^2 M_{\tau\tau} + 2s_{23} c_{23} M_{\mu\tau}.
\label{Eq:Parameters}
\end{eqnarray}
This formula reproduces the obtained results Eq.(\ref{Eq:mixing_13case}) for 
$c_{23}=\sigma s_{23}=1/\sqrt{2}$ and Eq.(\ref{Eq:mixing_12case}) for 
$c_{23}=-\sigma s_{23}=1/\sqrt{2}$ because $M_{sym}$ is specified by 
$M_{e\mu}=M^{(+)}_{e\mu}$, $M_{e\tau}=-\sigma M^{(+)}_{e\mu}$ and 
$M_{\mu\mu}=M_{\tau\tau}=M^{(+)}_{\mu\mu}$.

It is readily found that the predictions of masses from Eq.(\ref{Eq:ExactMasses})
 are identical to the three eigenvalues in each mass-ordering.  For instance, 
in the case of $\sin\theta_{12}=0$, we find that $m_1=\lambda_1$, which becomes
\begin{eqnarray}
&&
m_1 = M_{\mu\mu} - \sigma M_{\mu\tau} + \frac{1}{{1 - t_{13}^2 }}\left( {M_{ee} 
- M_{\mu\mu}+\sigma M_{\mu\tau} } \right).
\label{Eq:Example-12=0}
\end{eqnarray}
By using 
\begin{eqnarray}
&&
t_{13}  = \sigma \frac{{\sqrt {\left( {M_{ee}  - M_{\mu \mu }  + \sigma M_{\mu 
\tau } } \right)^2  + 8M_{e\mu }^2 }  - \left( {M_{ee}  - M_{\mu \mu }  + \sigma 
M_{\mu \tau } } \right)}}{{2\sqrt 2 M_{e\mu } }},
\label{Eq:t-13-squared}
\end{eqnarray}
calculated from $\tan 2\theta_{13}$ in  Eq.(\ref{Eq:ExactMixingAngles}), we reach
\begin{eqnarray}
m_1 &=& M_{\mu\mu} - \sigma M_{\mu\tau} 
\nonumber\\
&&
+ \frac{1}{2}\left( M_{ee} - M_{\mu\mu} + \sigma M_{\mu\tau}
+ \sqrt{ \left( {M_{ee} - M_{\mu\mu} + \sigma M_{\mu\tau}} \right)^2  +8M^2_{e\mu}
}  \right).
\label{Eq:Example-m1}
\end{eqnarray}
Similarly, we find that
\begin{eqnarray}
m_3 &=& M_{\mu\mu} - \sigma M_{\mu\tau} 
\nonumber\\
&&
 + \frac{1}{2}\left({{M_{ee} - M_{\mu\mu} + \sigma M_{\mu\tau}}}-\sqrt {\left( 
{M_{ee} - M_{\mu\mu} + \sigma M_{\mu\tau}} \right)^2  + 8M^2_{e\mu} }  \right).
\label{Eq:Example-m3}
\end{eqnarray}
These results, respectively, coincide with $\lambda_+$ and $\lambda_-$, which is 
the case of Eq.(\ref{Eq:U_PMNS_12case}).

\section{\label{sec:3}Approximately $\mu$-$\tau$ Symmetric Texture}
To discuss how the $\mu$-$\tau$ symmetry breaking term $M_b$ gives consistent 
predictions with the observed masses and mixings, we rely upon the formula provided 
by Eqs.(\ref{Eq:ExactMixingAngles}) and (\ref{Eq:ExactMasses}).  Since the 
experimentally allowed value of $\sin^2\theta_{13}={\mathcal{O}}(10^{-2})$ can 
describe the hierarchical ratio of $\Delta m^2_\odot/\Delta m^2_{atm}$, we retain 
terms of ${\mathcal{O}}(\sin^2\theta_{13}) $ in our calculations.  The $\mu$-$\tau$
 symmetry breaking effect is characterized by the parameter $\varepsilon$, which 
control $M^{(-)}_{e\mu}$ and $M^{(-)}_{\mu\mu}$.  Our mass matrix, then, takes 
the following form:
\begin{eqnarray}
M_\nu=
\left( {\begin{array}{*{20}c}  
 a & b & { - \sigma b}  \\  
 b & d & e  \\  
 { - \sigma b} & e & d  \\
\end{array}} \right) + \varepsilon \left( {\begin{array}{*{20}c}  
 0 & {b^\prime} & {\sigma b^\prime}  \\  
 {b^\prime} & {d^\prime} & 0  \\  
 {\sigma b^\prime} & 0 & { - d^\prime}  \\
\end{array}} \right).
\label{Eq:M_13-break}
\end{eqnarray}
This texture is almost the same as the one discussed in Ref.\cite{mu-tau-breaking}.  
However, constraints on the flavor masses are not well clarified, and the case 
corresponding to $\sin\theta_{12}=0$ is not discussed.  These two subjects are 
examined in detail by focusing on the flavor structure of $M_\nu$.  We discuss 
how different flavor structure yielding the same mass pattern results in different 
predictions.  The $\mu$-$\tau$ symmetry breaking generally induces the deviation 
of the atmospheric neutrino mixing from the maximal one as indicated by Eq.(
\ref{Eq:ExactMixingAngles}) for $\theta_{23}$ because of $M_{\mu\mu}\neq M_{\tau\tau}$
 and $s_{13}\neq 0$.  This deviation is parameterized by $\Delta$:
\begin{eqnarray}
c_{23} = \frac{{1 + \Delta }}{{\sqrt {2\left( {1 + \Delta ^2 } \right)} }},
\quad
s_{23} = \pm \sigma \frac{{1 - \Delta }}{{\sqrt {2\left( {1 + \Delta ^2 } \right)} }},
\label{Eq:AtmDeviation}
\end{eqnarray}
giving $\sin 2\theta _{23} = \pm\sigma(1-\Delta^2)/(1+\Delta^2)$ and 
$\cos 2\theta_{23} = 2\Delta/(1+\Delta^2)$.  The plus (minus) sign in front of 
$\sigma$ for $s_{23}$ specifies textures with $\sin\theta_{13}\rightarrow 0$ (
$\sin\theta_{12}\rightarrow 0$) as $\varepsilon \rightarrow 0$.

The masses and mixing angles are given by the following equations:
\begin{enumerate}
\item[{C1)}] with $\sin\theta_{13}\rightarrow 0$ as $\varepsilon\rightarrow 0$:
\begin{eqnarray}
&&
m_1  \approx \frac{a + d - \sigma e - \left( {d + \sigma e - a} \right)t_{13}^2  +2 
\left( \sigma e \Delta  +\varepsilon d^\prime  \right)\Delta }{2} - \frac{{X}}{{
\sin 2\theta _{12} }},
\nonumber \\
&&
m_2  \approx \frac{a + d - \sigma e - \left( {d + \sigma e - a} \right)t_{13}^2  +2 
\left(\sigma e\Delta +\varepsilon d^\prime \right)\Delta }{2} + \frac{{X}}{{\sin 
2\theta _{12} }},
\nonumber \\
&&
m_3  \approx d + \sigma e + \left( {d + \sigma e - a} \right)t_{13}^2  -2 \left
(  \sigma e \Delta  +  \varepsilon d^\prime \right)\Delta,
\label{Eq:Masses-13}
\end{eqnarray}
and
\begin{eqnarray}
&&
\tan 2\theta _{12}  \approx \frac{{2X}}{{d - \sigma e - a + \left( {d + \sigma e 
- a} \right)t_{13}^2  +2 \left( \sigma e\Delta  +\varepsilon d^\prime  \right)
\Delta }},
\nonumber \\
&&
\tan 2\theta _{13}  \approx \frac{{2Y}}{{d + \sigma e - a -2 \left( \sigma e\Delta  
+  \varepsilon d^\prime  \right)\Delta }},
\nonumber \\
&&
\cos 2\theta_{23} \approx 2\Delta,
\quad
\sin 2\theta _{23} \approx \sigma,
\label{Eq:Mixing-13}
\end{eqnarray}
with
\begin{eqnarray}
&&
X \approx \sqrt 2 \left( {b\left(1+\frac{t^2_{13}-\Delta^2}{2}\right) + \varepsilon 
b^\prime \Delta } \right),
\quad
Y \approx \sqrt 2 \sigma \left( {\varepsilon b^\prime  - b\Delta } \right),
\quad
\Delta  \approx  - \frac{{\sigma \varepsilon d^\prime + \sqrt 2 s_{13} b }}{{2e}},
\label{Eq:X-Y-Delta-13}
\end{eqnarray}
where we see the result of $\sin\theta_{13}\rightarrow 0$ as $\varepsilon\rightarrow 0$.

\item[{C2)}] with $\sin\theta_{12}\rightarrow 0$ as $\varepsilon\rightarrow 0$:
\begin{eqnarray}
&&
m_1  \approx \frac{{a + d + \sigma e - \left( {d - \sigma e - a} \right)t_{13}^2  
-2 \left( \sigma e\Delta  -\varepsilon d^\prime \right)\Delta }}{2} - \frac{X}{{
\sin 2\theta _{12} }},
\nonumber \\
&&
m_2  \approx \frac{{a + d + \sigma e - \left( {d - \sigma e - a} \right)t_{13}^2  
- 2\left( \sigma e\Delta  -  
\varepsilon d^\prime \right)\Delta }}{2} + \frac{X}{{\sin 2\theta _{12} }},
\nonumber \\
&&
m_3  \approx d - \sigma e + \left( {d - \sigma e - a} \right)t_{13}^2  + 2\left( 
\sigma e\Delta  - \varepsilon d^\prime \right)\Delta ,
\label{Eq:Masses-12}
\end{eqnarray}
and
\begin{eqnarray}
&&
\tan 2\theta _{12}  \approx \frac{{2X}}{{d + \sigma e - a + 2 \left( {d - \sigma 
e - a} \right)t_{13}^2  -2 \left(\sigma e \Delta  -  \varepsilon d^\prime \right)
\Delta }},
\nonumber \\
&&
\tan 2\theta _{13}  \approx \frac{{2Y}}{{d - \sigma e - a +2 \left( \sigma e\Delta  
-\varepsilon d^\prime  \right)\Delta }}
\nonumber \\
&&
\cos 2\theta_{23} \approx 2\Delta,
\quad
\sin 2\theta _{23} \approx -\sigma,
\label{Eq:Mixing-12}
\end{eqnarray}
with
\begin{eqnarray}
&&
X \approx \sqrt 2 \left( {\varepsilon b^\prime  + b\Delta } \right),
\quad
Y \approx  - \sqrt 2 \sigma \left( {b\left( 1-\frac{\Delta^2}{2}\right) - \varepsilon 
b^\prime \Delta } \right),
\quad
\Delta  \approx \frac{{\sigma d^\prime  - \sqrt 2 s_{13} b^\prime }}{{2e + \sqrt 
2 s_{13} b}}\varepsilon,
\label{Eq:X-Y-Delta-12}
\end{eqnarray}
where $\sin\theta_{12}\rightarrow 0$ as $\varepsilon\rightarrow 0$. It should be 
stressed again that the smallness of $\sin^2\theta_{13}$ is not guaranteed by the 
$\mu$-$\tau$ symmetry because $Y$ is mainly proportional to $b$, namely, to 
$M^{(+)}_{e\mu}$.  To obtain its smallness needs an additional requirement.
\end{enumerate}

It appears that, roughly speaking, masses given in C2) are almost the same as 
those in C1) by the change of $\sigma\rightarrow -\sigma$ for $e$ as $M_{\mu\tau}$.  
This correspondence occurs only if the contributions from $X$ are suppressed 
in C1) because the suppression factor $\varepsilon$ as shown in Eq.(\ref{Eq:X-Y-Delta-12})
 is always accompanied by those in C2).  However, $X$ needs not be suppressed 
in textures for C1), which cannot lead to textures in C2) by the change of 
$\sigma\rightarrow -\sigma$. 

\section{\label{sec:4}General Results}
In this section, we discuss general features present in C1) and C2) as a consequence 
of the tiny $\mu$-$\tau$ symmetry breaking, whose effects on the neutrino masses 
and mixings are evaluated in the previous section.   First of all, the relation 
between $\Delta m^2_\odot$ and $X$ can be expressed as 
\begin{eqnarray}
&&
\Delta m_ \odot ^2  = 
\frac{{2\sqrt 2 \left( {m_1  + m_2 } \right)X}}{{\sin 2\theta _{12} }}.
\label{Eq:Delta-X}
\end{eqnarray}
The condition of $\Delta m^2_\odot/\Delta m^2_{atm}\ll 1$ requires that either  
$m_1  + m_2$, or $X$ is suppressed.  These two options are used in C1) and C2) 
as follows:
\begin{enumerate}
\item[{C1)}] The suppression of $X$ is not a natural consequence, we may have 
$m_1+m_2\approx 0$ \cite{PlusMinusNu}.  If $m_1+ m_2\approx 0$, $X$ needs not be 
suppressed.  The requirement of $m_1+m_2\approx 0$ can be fulfilled if $a+d-\sigma e= 0$, 
more precisely, $\vert a+d-\sigma e\vert\ll\varepsilon^2$, is satisfied.  The 
sum of $m_1+m_2$ turns out to be ${\mathcal{O}}(\sin^2\theta_{13})$, which arises 
from the terms of $t^2_{13}$, $\varepsilon$ and $\Delta$ in Eq.(\ref{Eq:Masses-13})
.  As a result, we obtain that
\begin{eqnarray}
&&
\frac{\Delta m^2_\odot}{\Delta m^2_{atm}}\sim \sin^2\theta_{13},
\label{Eq:Delta-13}
\end{eqnarray}
which is one of the main results found in this article.   In the case of 
$m_1+m_2\neq 0$, $X$ should be suppressed and $b\approx 0$ is required because 
the $b^\prime$-term in $X$ is more suppressed by the factor $\varepsilon^2$.
\item[{C2)}] The suppression of $X$ is a consequence of the tiny $\mu$-$\tau$ 
symmetry breaking that leads to $X\propto\varepsilon$.  However, the phenomenological 
requirement of $\sin^2\theta_{13}\ll 1$, thereby, of the relative smallness of 
$Y$, must be satisfied.  From $Y \approx -\sqrt{2}\sigma (b-\varepsilon b^\prime \Delta )$, 
we may have $Y \approx -\sqrt{2}\sigma b$.  The suppression of $Y$ can be 
achieved by the smallness of $b$. The similar situation to Eq.(\ref{Eq:Delta-13})
 also arises in C2), but $\Delta m^2_\odot/\Delta m^2_{atm}$ receives an extra 
suppression due to $\varepsilon$ present in $X$, which makes $\Delta m^2_\odot/\Delta m^2_{atm}$
 phenomenologically unacceptable. 
\end{enumerate}
To obtain Eq.(\ref{Eq:Delta-13}), we can show even in our general discussions 
that the mass orderings of neutrinos are not arbitrary.  Since 
$m_2 \approx -m_1 \approx \sqrt{2} X/\sin 2\theta_{12}$ and $m_3\approx d+e\sigma$, 
the relation can be obtained for the quasi degenerate mass pattern since both 
$\vert d+e\sigma\vert$ and $\vert X/\sin 2\theta_{12}\vert$ are not suppressed 
and give $\vert m_{1,2,3}\vert \sim\sqrt{\Delta m^2_{atm}}$, or for the inverted 
mass hierarchy if $d+e\sigma\sim 0$, giving $\vert m_3\vert \sim\sin^2\theta_{13}\vert m_{1,2}\vert$
 \cite{New}.

It is instructive to note that the smallness of $b$ can be ascribed to the 
approximate conservation of the electron number $L_e$. Namely, $a$ has $L_e=2$, 
$b,b^\prime$ have $L_e=1$ and $d,e,d^\prime$ have $L_e=0$ \cite{eNumber-example,eNumber}.
 In the case that the conservation of $L_e$ is perturbatively violated by an 
interaction of $\vert\Delta L_e\vert = 1$ with an appropriate small parameter 
$\eta$ \cite{eNumber-example}, it is not absurd to expect $ a \propto \eta^2$, 
$ b \propto \eta$  and $d,e \propto \eta^0$, which explain the required suppression 
of $b$. This mechanism is only possible for the normal mass hierarchy.  It is 
because the condition of $\tan 2\theta_{12}={\mathcal{O}}(1)$ requires $d+\sigma e\sim 0$
 for $a\sim 0$, which gives $m_{1,2}\sim 0$. If this is the case, we obtain that
\begin{enumerate}
\item[{C1)}] $\sin^2\theta_{13}\ll 1$ due to the approximate $\mu$-$\tau$ symmetry 
and $\Delta m^2_\odot/\Delta m^2_{atm}\ll 1$ due to the approximate $L_e$-conservation 
(as long as $m_1+m_2\neq 0$).
\item[{C2)}] $\sin^2\theta_{13}\ll 1$ due to the approximate $L_e$-conservation 
and $\Delta m^2_\odot/\Delta m^2_{atm}\ll 1$ due to the approximate the $\mu$-
$\tau$ symmetry.
\end{enumerate}
Therefore, any underlying dynamics equipped with these two symmetries can describe 
the gross feature of the neutrino oscillations as the normal mass hierarchy.

There is a severe constraint on the flavor neutrino masses in C2) in order to 
satisfy $\sin^22\theta_{12}\sim {\mathcal{O}}(1)$.  Because $X$ receives the $\mu$
-$\tau$ symmetry breaking, $X$ is suppressed.  If $d + \sigma e - a\approx 0$, 
we have $\tan 2\theta_{12}\gg 1$, leading to the almost maximal mixing, since the 
corrections to $d + \sigma e - a$ in the denominator of $\tan 2\theta_{12}$ in 
Eq.(\ref{Eq:Mixing-12}) are ${\mathcal{O}}(\varepsilon^2)$.  We must obtain that 
$\vert d + \sigma e - a\vert\propto \vert\varepsilon\vert$, leading to
\begin{eqnarray}
&&
\tan 2\theta _{12}  \approx \frac{{2X}}{{d + \sigma e - a}}.
\label{Eq:Mixing12-finite}
\end{eqnarray}
As a result, the denominator cancels $\varepsilon$ in the numerator to yield 
$\tan 2\theta_{12}\sim {\mathcal{O}}(1)$.  Therefore, any textures realized in 
C2) must satisfy that
\begin{eqnarray}
&&
\vert d + \sigma e - a\vert \propto \vert\varepsilon\vert,
\label{Eq:Mixing12-finite^cond}
\end{eqnarray}
to match with $\sin^22\theta_{12}\sim {\mathcal{O}}(1)$. It means that the magnitude 
of $d + \sigma e - a$ should be adjusted so as to become as small as that of 
$\varepsilon$.

Our formula further shows a general relation between $\cos 2\theta_{23}$ and other 
small quantities \cite{Theta31AndMass} arising from the effect of the tiny $\mu$
-$\tau$ symmetry breaking.  Such a relation comes from the formula of Eq.(
\ref{Eq:ExactMixingAngles}) for $\theta_{23}$ and is readily found that
\begin{enumerate}
\item[{C1)}] because $M^{(-)}_{\mu\mu}$ and $s_{13}$ receive the $\mu$-$\tau$ 
symmetry breaking effect, their sizes are proportional to $\varepsilon$, leading 
to
\begin{eqnarray}
&&
\cos 2\theta_{23}\approx \frac{\sigma M^{(-)}_{\mu\mu}-s_{13}X}{M_{\mu\tau}}\propto 
\varepsilon \sim \sin\theta_{13},
\label{Eq:cos23-13}
\end{eqnarray}
and
\item[{C2)}] because $X$ receives the $\mu$-$\tau$ symmetry breaking effect while 
$s_{13}$ is required to be phenomenologically suppressed, the product of $Xs_{13}$
 is doubly suppressed and can be a vanishingly small quantity, leading to
\begin{eqnarray}
&&
\cos 2\theta_{23}\sim \frac{\sigma M^{(-)}_{\mu\mu}}{M_{\mu\tau}}\propto \varepsilon 
\sim \frac{\Delta m^2_\odot}{\Delta m^2_{atm}},
\label{Eq:cos23-12}
\end{eqnarray}
where $\varepsilon$ can be related to $\Delta m^2_\odot/\Delta m^2_{atm}$ because 
$X\propto \varepsilon$ in Eq.(\ref{Eq:Delta-X}).
\end{enumerate}
The faithful parameter measuring the size of the $\mu$-$\tau$ symmetry breaking 
effect is $\cos 2\theta_{23}$.

 From Eqs.(\ref{Eq:X-Y-Delta-13}) for C1) and (\ref{Eq:X-Y-Delta-12}) for C2), 
we can, respectively, obtain that
\begin{eqnarray}
&&
s_{13}  \approx \frac{{2 eb^\prime  + \sigma bd^\prime }}{{\sqrt 2 \left[ {\sigma 
e\left( {d + \sigma e - a} \right) - b^2 } \right]}}\varepsilon,
\quad
\cos 2\theta _{23}  \approx  - \frac{{ \left( {d + \sigma e - a} \right)d^\prime }
 + 2 bb^\prime}{{\sigma e\left( {d + \sigma e - a} \right) - b^2 }}\varepsilon,
\label{Eq:s13Delta-13}
\end{eqnarray}
for $d + \sigma e - a\neq 0$, where $s_{13}$ and $\cos 2\theta_{23}$ satisfy Eq.
(\ref{Eq:cos23-13}), and
\begin{eqnarray}
&&
s_{13}  \approx  - \frac{\sqrt 2 \sigma b}{d - \sigma e - a},
\quad
\cos 2\theta _{23}  \approx \frac{\left( d - \sigma e - a \right)d^\prime  + 2bb^
\prime }{\sigma e\left( {d - \sigma e - a} \right) - b^2}\varepsilon,
\label{Eq:s12Delta-13}
\end{eqnarray}
for $d - \sigma e - a\neq 0$, which coincides with $\cos 2\theta _{23}$ of Eq.(
\ref{Eq:cos23-12}) if the phenomenological requirement of $\sin^2\theta_{13}\ll 1$
 is translated into the smallness of $b$.

\section{\label{sec:5}Summary and Discussions}
We have clarified the effects from the $\mu$-$\tau$ symmetry breaking in neutrino 
mass textures.  It is of great significance to recognize that the ordering of the 
eigenvalues for a given neutrino mass matrix conceptually yields completely 
different results.  If the texture is $\mu$-$\tau$ symmetric, its diagonalization 
gives either $\sin\theta_{13}=0$ for C1) or $\sin\theta_{12}=0$ for C2).  Of 
course, the case of $\sin\theta_{13}=0$ is a usually claimed result if the $\mu$
-$\tau$ symmetry is present.  However, the case with $\sin\theta_{12}=0$ is equally 
possible to arises and the $\mu$-$\tau$ symmetry breaking points to the suppression 
of $\Delta m^2_\odot/\Delta m^2_{atm}$.  Practically, including the $\mu$-$\tau$
 symmetry breaking effect of $\varepsilon$, we can choose the case of $\sin\theta_{12}=0$
 as a phenomenologically acceptable one once a fine-tuning is invoked to yield 
$\sin^2\theta_{12}={\mathcal{O}}(1)$ provided that another requirement of 
$\sin^2\theta_{13}\ll 1$ is fulfilled.  This observation indicates that the C2) 
case necessarily involves two small quantities, which is $\varepsilon$, and another 
quantity $\eta$ that keeps $\sin^2\theta_{13}\ll 1$.  However, it would be curious 
to have $\sin\theta_{12}=0$ in the symmetric limit. It is only possible if the 
zero-th order contribution is as small as $\varepsilon$.  Since the $\varepsilon$
-term is placed on the numerator of $\tan 2\theta_{12}$, the necessary condition 
to have $\sin^22\theta_{12}={\mathcal{O}}(1)$ is to require $\vert d + \sigma e - a\vert \sim \eta$
 with $\vert\eta\vert\sim \vert\varepsilon\vert$ in the denominator.

The relations among $\cos 2\theta_{23}$, $\sin\theta_{13}$ and 
$\Delta m^2_\odot/\Delta m^2_{atm}$ are shown to indicate the general property 
for any texture as described in Eqs.(\ref{Eq:cos23-13}) and (\ref{Eq:cos23-12}).  
We have found the relations:
\begin{eqnarray}
&&
\cos 2\theta_{23}\sim \sin\theta_{13},
\quad
\cos 2\theta_{23}\sim \frac{\Delta m^2_\odot}{\Delta m^2_{atm}},
\label{Eq:new-relations-summary}
\end{eqnarray}
respectively, for C1) and C2).  It is clear that there is no relationship between 
$\sin\theta_{13}$ and $\Delta m^2_\odot/\Delta m^2_{atm}$ as long as the results 
of the $\mu$-$\tau$ symmetry breaking are concerned.  To have any correlation 
between $\sin^2\theta_{13}$ and $\Delta m^2_\odot/\Delta m^2_{atm}$, we need some 
specific relations among the flavor masses, which currently arise from other 
phenomenological requirements.

We have also argued that the suppression of $\Delta m^2_\odot$, which is estimated 
to be $\Delta m_ \odot ^2  = 2\sqrt 2 (m_1  + m_2  )X/\sin 2\theta _{12}$, requires 
either 
\begin{itemize}
\item $m_1+m_2\approx 0$, or
\item $X\approx 0$, where $\tan 2\theta_{12}$ is proportional to $X$.
\end{itemize}
To satisfy $X\approx 0$ is a consequence of the approximate $\mu$-$\tau$ symmetry 
breaking in C2).  On the other hand, the C1) case requires $m_1+m_2\approx 0$ 
including the suppression of both $m_1$ and $m_2$ such as in the normal mass 
hierarchy.  If we demand that $a+d-\sigma e= 0$, we obtain $m_1+ m_2\propto \sin^2\theta_{13}$
 from Eq.(\ref{Eq:Masses-13}), leading to
\begin{eqnarray}
&&
\frac{\Delta m^2_\odot}{\Delta m^2_{atm}}\sim \sin^2\theta_{13},
\label{Eq:MassSquared-summary}
\end{eqnarray}
as in Eq.(\ref{Eq:Delta-13}), which arises from contributions of 
${\mathcal{O}}(\sin^2\theta_{13})$.  It is also argued that this relation is only 
possible for the inverted mass hierarchy, and the quasi degenerate mass pattern 
with $\vert m_{1,2,3}\vert\sim\sqrt{\Delta m^2_{atm}}$.  The similar relation 
also exists for C2), but it may not be phenomenologically acceptable because of 
the further suppression of $\Delta m^2_\odot/\Delta m^2_{atm}$ due to $X$.

Additional necessary suppression is required in the C2) case to meet 
$\sin^2\theta_{13}\ll 1$, which may be due to the approximate $L_e$ conservation.  
The presence of this conservation also helpful to have $\Delta m^2_\odot\ll 1$
 for  C1) with  $m_1+m_2\neq 0$.  We expect the following scenarios to emerge.  
First, any underlying dynamics equipped with the approximate $\mu$-$\tau$ symmetry,
 and the approximate $L_e$-conservation explains $\sin^2\theta_{13}\ll 1$ as well 
as $\Delta m^2_\odot/\Delta m^2_{atm}\ll 1$.  Furthermore, in C2), 
$\sin^22\theta_{12}={\mathcal{O}}(1)$ should be realized and can be obtained if 
the dynamics ensures that $\vert d + \sigma e - a\vert \propto \vert\varepsilon\vert$.  
Next, especially in C1), the dynamics only equipped with the approximate $\mu$
-$\tau$ symmetry can describe $\sin^2\theta_{13}\ll 1$ and $\Delta m^2_\odot/\Delta m^2_{atm}\ll 1$
 if it ensures that $a+d-\sigma e = 0$.

In conclusion, we have demonstrated how the argument based on the $\mu$-$\tau$ 
symmetry is powerful not only for the classification of textures but also for the 
discovery of general correlations among $\sin\theta_{13}$, $\cos 2\theta_{23}$ 
and $\Delta m^2_\odot/\Delta m^2_{atm}$.  The origin of $\Delta m^2_\odot/\Delta m^2_{atm}\ll 1$
 can be ascribed to the approximate $L_e$ conservation in the normal mass hierarchy,
 and to the relationship of $\Delta m^2_\odot/\Delta m^2_{atm}\sim\sin^2\theta_{13}$
 in the inverted mass hierarchy, and the quasi degenerate mass pattern.  More 
precise estimation of these relations is possible if we construct explicit textures 
leading to the normal and inverted mass hierarchies, and to the quasi degenerate 
mass pattern. 

\vspace{3mm}
\noindent
\centerline{\small \bf ACKNOWLEGMENTS}

The authors are grateful to I. Aizawa and T. Kitabayashi for useful discussions.  
The work of M.Y. is supported by the Grants-in-Aid for Scientific Research on 
Priority Areas (No 13135219) from the Ministry of Education, Culture, Sports, 
Science, and Technology, Japan.





\end{document}